\documentclass{article}
\usepackage{amsmath}
\usepackage{amssymb}

\input{tcilatex}
\begin{document}

\begin{center}
463SolvDiscrTimeDynSyst190203\bigskip

{\Huge Solvable Systems Featuring 2 Dependent Variables Evolving in
Discrete-Time via 2 Nonlinearly-Coupled First-Order Recursion Relations with
Polynomial Right-Hand Sides}

\textbf{Francesco Calogero}$^{a,b,1}${\LARGE \ {\large and} }\textbf{Farrin
Payandeh}$^{a,c,2}$

$^{a}$ Physics Department, University of Rome "La Sapienza", Rome, Italy

$^{b}$ INFN, Sezione di Roma 1

$^{c}$ Department of Physics, Payame Noor University (PNU), PO BOX
19395-3697 Tehran, Iran

$^{1}$ francesco.calogero@roma1.infn.it, francesco.calogero@uniroma1.it

$^{2}$ f\_payandeh@pnu.ac.ir, farrinpayandeh@yahoo.com

\bigskip

\textit{Abstract}
\end{center}

The evolution equations mentioned in the title of this paper read as follows:%
\begin{equation*}
\tilde{x}_{n}=P^{\left( n\right) }\left( x_{1},x_{2}\right) ~,~~~n=1,2~,
\end{equation*}%
where $\ell $ is the "discrete-time" \textit{independent} variable taking
\textit{integer} values ($\ell =0,1,2,...$), $x_{n}\equiv x_{n}\left( \ell
\right) $ are the $2$ dependent variables, $\tilde{x}_{n}\equiv x_{n}\left(
\ell +1\right) $, and the $2$ functions $P^{\left( n\right) }\left(
x_{1},x_{2}\right) ,~n=1,2,$ are $2$ \textit{polynomials} in the $2$
dependent variables $x_{1}\left( \ell \right) $ and $x_{2}\left( \ell
\right) $. The results reported in this paper have been obtained by an
appropriate modification of a recently introduced technique to obtain
analogous results in \textit{continuous-time} $t$---in which case $%
x_{n}\equiv x_{n}\left( t\right) $ and the above recursion relations are
replaced by first-order ODEs. Their potential interest is due to the
relevance of this kind of evolution equations in various applicative
contexts.

\bigskip

\section{Introduction}

In this introductory \textbf{Section 1}, after providing some notational
prescriptions, we tersely review previous relevant findings.

\textbf{Notation 1-1}. Hereafter $\ell =0,1,2,...$denotes the \textit{%
discrete-time} \textit{independent} variable; the \textit{dependent}
variables are $x_{n}\equiv x_{n}\left( \ell \right) $ (generally with $n=1,2$%
), and the notation $\tilde{x}_{n}\equiv x_{n}\left( \ell +1\right) $
indicates the once-updated values of these variables. We shall also use
other dependent variables, for instance $y_{m}\equiv y_{m}\left( \ell
\right) ,$ and then of course likewise $\tilde{y}_{m}\equiv y_{m}\left( \ell
+1\right) $. All variables such as $x$, $y$, $z$, (generally equipped with
indices) are assumed to be \textit{complex} numbers, unless otherwise
indicated; it shall generally be clear from the context which of these and
other quantities depend on time (as occasionally---but not always---\textit{%
explicitly} indicated); parameters such as $a$, $\alpha ,$ $\beta ,$ $\gamma
,$ $A,$ etc. (often equipped with indices) are generally time-independent
\textit{complex} numbers; and indices such as $n$, $m$, $j$ are generally
\textit{positive integers} (the values they may take shall be explicitly
indicated or quite clear from the context). $\blacksquare $

\textbf{Remark 1-1}. In this paper the term \textit{solvable} generally
characterizes systems of evolution equations the initial-values problems of
which are \textit{explicitly solvable by algebraic operations}. $%
\blacksquare $

In the following \textbf{Subsection 1.1} we tersely review---mainly via
quotations (with minor adjustments) from a recent paper of ours \cite{CP2019}%
---a recent approach to identify \textit{solvable} dynamical systems in
\textit{continuous-time} $t$, as introduction to the extension of (some of)
these results to the case of \textit{discrete-time} $\ell $, which is the
topic of the present paper. Previous results on \textit{solvable
discrete-time} models are tersely reviewed in the subsequent \textbf{%
Subsection 1.2}. Our main findings are reported in \textbf{Section 2} (also
based on the results reported in \textbf{Appendix A}). A concluding \textbf{%
Section 3} outlines tersely possible additional developments.

\bigskip

\subsection{Review of an analogous approach in the \textit{continuous-time}
context}

"Long time ago the idea has been introduced to identify dynamical systems
(evolving in \textit{continuous-time} $t$) which are \textit{solvable} by
using as a tool the relations between the time evolutions of the \textit{%
coefficients} and the \textit{zeros} of a generic time-dependent polynomial
\cite{C1978}. The basic idea of this approach is to relate the
time-evolution of the $N$ zeros $x_{n}\left( t\right) $ of a generic
time-dependent polynomial $p_{N}\left( z;t\right) $ of degree $N$ in its
argument $z,$%
\begin{subequations}
\label{pNzt}
\begin{equation}
p_{N}\left( z;t\right) =z^{N}+\sum_{m=1}^{N}\left[ y_{m}\left( t\right)
z^{N-m}\right] =\prod_{n=1}^{N}\left[ z-x_{n}\left( t\right) \right] ~,
\label{1pNzt}
\end{equation}%
to the time-evolution of its $N$ coefficients $y_{m}\left( t\right) $.
Indeed, if the time evolution of the $N$ coefficients $y_{m}\left( t\right) $
is determined by a system of ODEs which is itself \textit{solvable}, then
the corresponding time-evolution of the $N$ zeros $x_{n}\left( t\right) $ is
also \textit{solvable}, via the following $3$ steps: (i) given the initial
values $x_{n}\left( 0\right) ,$ the corresponding initial values $%
y_{m}\left( 0\right) $ can be obtained from the \textit{explicit}
formulas---expressing the $N$ coefficients $y_{m}\left( t\right) $ of the
polynomial (\ref{1pNzt}) in terms of its $N$ zeros $x_{n}\left( t\right) $%
---reading (for all time, hence in particular at $t=0$)%
\begin{equation}
y_{m}\left( t\right) =\left( -1\right) ^{m}\sum_{1\leq
n_{1}<n_{2}<...<n_{m}\leq N}^{N}\left\{ \prod_{\ell =1}^{M}\left[ x_{n_{\ell
}}\left( t\right) \right] \right\} ~,~~~m=1,2,...,N~;  \label{1ymxn}
\end{equation}%
(ii) from the $N$ values $y_{m}\left( 0\right) $ thereby obtained, the $N$
values $y_{m}\left( t\right) $ are then evaluated via the---assumedly
\textit{solvable}---system of ODEs satisfied by the $N$ coefficients $%
y_{m}\left( t\right) $; (iii) the $N$ values $x_{n}\left( t\right) $---i.
e., the $N$ solutions of the dynamical system satisfied by the $N$ variables
$x_{n}\left( t\right) $---are then determined as the $N$ zeros of the
polynomial, see (\ref{1pNzt}), itself known at time $t$ in terms of its $N$
coefficients $y_{m}\left( t\right) $ (the computation of the zeros of a
known polynomial being an \textit{algebraic} operation; of course generally
explicitly performable only for polynomials of degree $N\leq 4$)...

The viability of this technique to identify \textit{solvable} dynamical
systems depends of course on the availability of an \textit{explicit} method
to relate the time-evolution of the $N$ zeros of a \textit{polynomial} to
the corresponding time-evolution of its $N$ \textit{coefficients}. Such a
method was indeed provided in \cite{C1978}, opening the way to the
identification of a vast class of \textit{algebraically solvable} dynamical
systems (see also, for instance, \cite{C2001} and references therein); but
that approach was essentially restricted to the consideration of \textit{%
linear} time evolutions of the coefficients $y_{m}\left( t\right) $.

A development allowing to lift this quite strong restriction emerged
relatively recently \cite{C2016}, by noticing the validity of the \textit{%
identity}
\end{subequations}
\begin{equation}
\dot{x}_{n}=-\left[ \prod_{\ell =1,~\ell \neq n}^{N}\left( x_{n}-x_{\ell
}\right) \right] ^{-1}\sum_{m=1}^{N}\left[ \dot{y}_{m}\left( x_{n}\right)
^{N-m}\right]  \label{1xndot}
\end{equation}%
which provides a convenient \textit{explicit} relationship among the time
evolutions of the $N$ \textit{zeros} $x_{n}\left( t\right) $ and the $N$
\textit{coefficients} $y_{m}\left( t\right) $ of the generic polynomial (\ref%
{1pNzt}). This allowed a major enlargement of the class of \textit{%
algebraically solvable} dynamical systems identifiable via this approach:
for many examples see \cite{C2018} and references therein...

A new twist of this approach was then provided by its extension to \textit{%
nongeneric} polynomials featuring---for \textit{all} time---\textit{multiple}
zeros. The first step in this direction focussed on time-dependent
polynomials featuring for \textit{all} time a \textit{single double zero}
\cite{BC2018}; and subsequently significant progress has been made to treat
the case of polynomials featuring a \textit{single zero} of \textit{%
arbitrary multiplicity} \cite{B2018}. A convenient method was then provided
which is suitable to treat the most general case of polynomials featuring an
\textit{arbitrary} number of \textit{zeros} each of which features an
\textit{arbitrary multiplicity}. While all these developments might appear
to mimic scholastic exercises analogous to the discussion among medieval
scholars of how many angels might dance simultaneously on the tip of\ a
needle, they do indeed provide \textit{new tools} to identify \textit{new}
dynamical systems featuring interesting time evolutions (including systems
displaying remarkable behaviors such as \textit{isochrony} or \textit{%
asymptotic isochrony}: see for instance \cite{BC2018} \cite{B2018});
dynamical systems which---besides their intrinsic mathematical
interest---are quite likely to play significant roles in applicative
contexts...

We then focused on another twist of this approach to identify new \textit{%
solvable} dynamical systems which was introduced quite recently \cite{CP2018}%
. It is again based on the relations among the time-evolution of the \textit{%
coefficients} and the \textit{zeros} of time-dependent polynomials \cite%
{C2016} \cite{C2018} with \textit{multiple roots} (see \cite{BC2018}, \cite%
{B2018} and above); restricting moreover attention to such polynomials
featuring \textit{only }$2$ \textit{zeros}. Again, this might seem such a
strong limitation to justify the doubt that the results thereby obtained be
of much interest. But the effect of this restriction is to open the
possibility to identify \textit{algebraically solvable }dynamical models
characterized by the following systems of $2$ ODEs,%
\begin{equation}
\dot{x}_{n}=P^{\left( n\right) }\left( x_{1},x_{2}\right) ~,~~~n=1,2~,
\label{1xndotPol}
\end{equation}%
with $P^{\left( n\right) }\left( x_{1},x_{2}\right) $ $2$ \textit{polynomials%
} in the $2$ dependent variables $x_{1}\left( t\right) $ and $x_{2}\left(
t\right) $; hence systems of considerable interest, both from a theoretical
and an applicative point of view (see \cite{CP2018} and references quoted
there)." \cite{CP2019}

This completes our review---via a long quotation from a previous paper---of
recent developments concerning certain classes of standard dynamical systems
in \textit{continuous time}. In the present paper---after tersely reviewing,
in the following \textbf{Subsection 1.2}, some past results in the \textit{%
discrete-time} context---we focus on the derivation in such a context of
analogous results to some of those reported in the \textit{continuous-time}
context in \cite{CP2019}.

\bigskip

\subsection{Review of somewhat analogous past findings in the \textit{%
discrete-time} context}

Somewhat analogous results to those reviewed in the \textit{first} part of
the previous \textbf{Subsection 1.1 }have been developed over time in the
context of \textit{discrete-time} evolutions, by focussing on the evolution
of the \textit{zeros} of generic monic polynomials the \textit{coefficients}
of which evolve in a \textit{solvable} manner in \textit{discrete time}.

The new results reported below consists essential of extensions to the
\textit{discrete-time} context of the results outlined in the \textit{second}
part of the preceding \textbf{Subsection 1.1}. Note however that here and
below we actually dispense from a general discussion of the evolution of the
\textit{zeros} of a polynomial the \textit{coefficients} of which evolve in
\textit{discrete time} in a \textit{solvable} manner, both in the case of
\textit{generic} monic polynomials (as treated in Chapter 7 of \cite{C2018})
and in the case of the special polynomials of higher degree than $2$ which
nevertheless feature for all time only $2$ (of course \textit{multiple})
\textit{zeros }(as treated in \cite{CP2018} \cite{CP2019}); below we rather
employ the simpler technique---described in the following \textbf{Section 2}%
---to identify \textit{solvable} nonlinear evolution equations that emerged
from that approach and which actually subtends most of the \textit{explicit}
findings reported in \cite{CP2019}. Hence from the previous findings for
\textit{discrete-time} evolutions---see \cite{BC2017} and \textbf{Chapter 7}
("Discrete time") of \cite{C2018}---we only use below the following \textit{%
discrete-time} equivalent of the identity (\ref{1xndot}) (originating from
the polynomial (\ref{pNzt}) with $t$ replaced by $\ell $),%
\begin{equation}
\prod\limits_{j=1}^{N}\left( \tilde{x}_{n}-x_{j}\right) +\sum_{m=1}^{N}\left[
\left( \tilde{y}_{m}-y_{m}\right) \left( \tilde{x}_{n}\right) ^{N-m}\right]
=0~,
\end{equation}%
hence, for the $N=2$ case,
\begin{subequations}
\label{1xntildeN2}
\begin{equation}
\left( \tilde{x}_{n}-x_{1}\right) \left( \tilde{x}_{n}-x_{2}\right) +\left(
\tilde{y}_{1}-y_{1}\right) \tilde{x}_{n}+\tilde{y}_{2}-y_{2}=0~,~~~n=1,2
\label{1xntildeN2a}
\end{equation}%
of course with (see (\ref{1ymxn}))
\begin{equation}
y_{1}\left( \ell \right) =-\left[ x_{1}\left( \ell \right) +x_{2}\left( \ell
\right) \right] ~,~~~y_{2}\left( \ell \right) =x_{1}\left( \ell \right)
x_{2}\left( \ell \right) ~.  \label{1xntildeN2b}
\end{equation}

\bigskip

\section{A \textit{solvable }system of 2 nonlinearly coupled evolution
equations in \textit{discrete-time} satisfied by 2 dependent variables}

In this \textbf{Section 2} we present our main results, consisting in the
identification of a \textit{solvable} systems of $2$ nonlinearly-coupled
\textit{discrete-time} evolution equations belonging to the class
\end{subequations}
\begin{equation}
\tilde{x}_{n}=P^{\left( n\right) }\left( x_{1},x_{2}\right) ~,~~~n=1,2~,
\label{2xntilde}
\end{equation}%
where the $2$ functions $P^{\left( n\right) }\left( x_{1},x_{2}\right) $ are
$2$ appropriately identified \textit{polynomials} in the variables $%
x_{1}\left( \ell \right) $ and $x_{2}\left( \ell \right) $; a system which
is clearly the natural generalization to \textit{discrete time} of the
\textit{continuous-time} system (\ref{1xndotPol}). In particular we
demonstrate the \textit{solvable} character of the following dynamical
system:%
\begin{equation}
\tilde{z}_{n}=a_{n1}\left( z_{1}\right) ^{2}+a_{n2}\left( z_{2}\right)
^{2}+a_{n3}z_{1}z_{2}~,~~~n=1,2~,  \label{2zntilde}
\end{equation}%
with the $6$ parameters $a_{nj}$ ($n=1,2,$ $j=1,2,3$) explicitly given by $6$
algebraic expressions in terms of $6$ \textit{arbitrary} parameters.

Let the $2$ dependent variables $y_{1}\left( \ell \right) $ and $y_{2}\left(
\ell \right) $ evolve in discrete time according to the following \textit{%
discrete-time} evolution equations (the \textit{solvability} of which is
demonstrated in \textbf{Appendix A}):%
\begin{equation}
\tilde{y}_{1}=\alpha \left( y_{1}\right) ^{2}~,~~~\tilde{y}_{2}=\beta
^{2}\left( y_{1}\right) ^{2}y_{2}+\gamma \left( y_{1}\right) ^{4}~;
\label{21y12tilde}
\end{equation}%
and assume again that the $2$ variables $y_{1}\left( \ell \right) $ and $%
y_{2}\left( \ell \right) $ are related to the $2$ variables $x_{1}\left(
\ell \right) $ and $x_{2}\left( \ell \right) $ as follows (see (\ref%
{1xntildeN2b})):
\begin{equation}
y_{1}\left( \ell \right) =-\left[ x_{1}\left( \ell \right) +x_{2}\left( \ell
\right) \right] ~,~~~y_{2}\left( \ell \right) =x_{1}\left( \ell \right)
x_{2}\left( \ell \right) ~.  \label{21yyxx}
\end{equation}

\textbf{Remark 2-1}. Let us re-emphasize that, if the \textit{discrete-time}
evolution of the $2$ variables $y_{1}\left( \ell \right) $ and $y_{2}\left(
\ell \right) $ is \textit{solvable}, then the \textit{discrete-time}
evolution of the $2$ variables $x_{1}\left( \ell \right) $ and $x_{2}\left(
\ell \right) $ is as well \textit{solvable}, because the \textit{ansatz }(%
\ref{21yyxx}) can be inverted via an \textit{algebraic} operation, indeed
quite \textit{explicitly}, since it clearly implies that $x_{1}\left( \ell
\right) $ and $x_{2}\left( \ell \right) $ are the $2$ roots of the following
monic polynomial of degree $2$:
\begin{equation}
p_{2}\left( z\right) =z^{2}+y_{1}z+y_{2}=\left( z-x_{1}\right) \left(
z-x_{2}\right) ~.~~~\blacksquare
\end{equation}%
~

It is then a matter of trivial algebra---either via the formulas (\ref%
{1xntildeN2}) or directly from (\ref{21y12tilde}) and (\ref{21yyxx})---to
derive the following system of two \textit{discrete-time} evolution
equations satisfied by the $2$ dependent variables $x_{1}\left( \ell \right)
$ and $x_{2}\left( \ell \right) $:
\begin{subequations}
\begin{equation}
\left( \tilde{x}_{n}\right) ^{2}+\alpha \left( x_{1}+x_{2}\right) ^{2}\tilde{%
x}_{n}+\left( x_{1}+x_{2}\right) ^{2}\left[ \beta ^{2}x_{1}x_{2}+\gamma
\left( x_{1}+x_{2}\right) ^{2}\right] =0~,
\end{equation}%
implying%
\begin{equation}
\tilde{x}_{n}=-\frac{\alpha \left( x_{1}+x_{2}\right) ^{2}+\left( -1\right)
^{n}\Delta }{2}~,
\end{equation}%
with
\begin{equation}
\Delta ^{2}=\left( x_{1}+x_{2}\right) ^{2}\left[ \left( \alpha ^{2}-4\gamma
\right) \left( x_{1}+x_{2}\right) ^{2}-4\beta ^{2}x_{1}x_{2}\right] ~.
\label{2Delta}
\end{equation}

Assume now that
\end{subequations}
\begin{subequations}
\begin{equation}
\gamma =\frac{\alpha ^{2}-\beta ^{2}}{4}~,  \label{21gamma}
\end{equation}%
so that the right-hand side of (\ref{2Delta}) become an \textit{exact}
square implying%
\begin{equation}
\Delta =\pm \beta \left[ \left( x_{1}\right) ^{2}-\left( x_{2}\right) ^{2}%
\right] ~.
\end{equation}%
Then clearly
\end{subequations}
\begin{subequations}
\begin{equation}
\tilde{x}_{n}=-\frac{\alpha \left( x_{1}+x_{2}\right) ^{2}+\left( -1\right)
^{n}\beta \left[ \left( x_{1}\right) ^{2}-\left( x_{2}\right) ^{2}\right] }{2%
}~,~~~n=1,2~,
\end{equation}%
namely%
\begin{equation}
\tilde{x}_{n}=a_{1}^{\left( n\right) }\left( x_{1}\right) ^{2}+a_{2}^{\left(
n\right) }\left( x_{2}\right) ^{2}+a_{3}^{\left( n\right)
}x_{1}x_{2}~,~~~n=1,2~,
\end{equation}%
with%
\begin{equation}
a_{1}^{\left( n\right) }=-\frac{\alpha +\left( -1\right) ^{n}\beta }{2}%
~,~~~a_{2}^{\left( n\right) }=-\frac{\alpha +\left( -1\right) ^{n}\beta }{2}%
~,~~~a_{3}^{\left( n\right) }=-\alpha ~.  \label{2a123n}
\end{equation}

We have thereby identified a simple \textit{solvable} system of type (\ref%
{2xntilde}), featuring in its right-hand side $2$ \textit{homogeneous}
polynomials of \textit{second} degree in the $2$ variables $x_{1}\left( \ell
\right) $ and $x_{2}\left( \ell \right) $, the $6$ coefficients $%
a_{j}^{\left( n\right) }$ ($n=1,2,~~j=1,2,3$) of which depend on the $2$
\textit{a priori arbitrary} parameters $\alpha $ and $\beta $, see (\ref%
{2a123n}).

From this system an analogous system featuring more arbitrary parameters can
be identified via the simple trick of introducing $2$ new dependent
variables, $z_{1}\left( \ell \right) $ and $z_{2}\left( \ell \right) ,$
related linearly to the $2$ variables $x_{1}\left( \ell \right) $ and $%
x_{2}\left( \ell \right) $:
\end{subequations}
\begin{subequations}
\begin{equation}
z_{1}=A_{11}x_{1}+A_{12}x_{2}~,~~~z_{2}=A_{21}x_{1}+A_{22}x_{2}~,
\end{equation}%
\begin{equation}
x_{1}=\left( A_{22}z_{1}-A_{12}z_{2}\right) /D~,~~~x_{2}=\left(
-A_{21}z_{1}+A_{11}z_{2}\right) /D~,
\end{equation}
\begin{equation}
D=A_{11}A_{22}-A_{12}A_{21}~.
\end{equation}%
It is easily seen that the new system is then just the system (\ref{2zntilde}%
), with the $6$ parameters $a_{nj}$ ($n=1,2,$ $j=1,2,3$) \textit{explicitly}
expressed as follows in terms of the $4$ \textit{arbitrary} parameters $%
A_{nm}$\ ($n=1,2,$ $m=1,2$) and the $2$ \textit{arbitrary} parameters $%
\alpha $ and $\beta $ (see (\ref{2a123n})):
\end{subequations}
\begin{subequations}
\begin{eqnarray}
a_{n1} &=&D^{-2}\left[ \left( A_{22}\right) ^{2}\left( A_{n1}a_{1}^{\left(
1\right) }+A_{n2}a_{1}^{\left( 2\right) }\right) +\left( A_{21}\right)
^{2}\left( A_{n1}a_{2}^{\left( 1\right) }+A_{n2}a_{2}^{\left( 2\right)
}\right) \right.  \notag \\
&&\left. -A_{22}A_{21}\left( A_{n1}a_{3}^{\left( 1\right)
}+A_{n2}a_{3}^{\left( 2\right) }\right) \right] ~,~~~n=1,2~,
\end{eqnarray}

\begin{eqnarray}
a_{n2} &=&D^{-2}\left[ \left( A_{12}\right) ^{2}\left( A_{n1}a_{1}^{\left(
1\right) }+A_{n2}a_{1}^{\left( 2\right) }\right) +\left( A_{11}\right)
^{2}\left( A_{n1}a_{2}^{\left( 1\right) }+A_{n2}a_{2}^{\left( 2\right)
}\right) \right.  \notag \\
&&\left. -A_{11}A_{12}\left( A_{n1}a_{3}^{\left( 1\right)
}+A_{n2}a_{3}^{\left( 2\right) }\right) \right] ~,~~~n=1,2~,
\end{eqnarray}%
\begin{eqnarray}
a_{n3} &=&D^{-2}\left[ -2A_{12}A_{22}\left( A_{n1}a_{1}^{\left( 1\right)
}+A_{n2}a_{1}^{\left( 2\right) }\right) -2A_{21}A_{11}\left(
A_{n1}a_{2}^{\left( 1\right) }+A_{n2}a_{2}^{\left( 2\right) }\right) \right.
\notag \\
&&\left. +\left( A_{11}A_{22}+A_{12}A_{21}\right) \left( A_{n1}a_{3}^{\left(
1\right) }+A_{n2}a_{3}^{\left( 2\right) }\right) \right] ~,~~~n=1,2~.
\end{eqnarray}

For the inversion of these transformations---i. e., the issue of expressing
the $4$ parameters $A_{nm}$ ($n=1,2;$ $m=1,2$) and the $2$ parameters $%
\alpha ,$ $\beta $ (see (\ref{2a123n})) in terms of the $6$ parameters $%
a_{n\ell }$ ($n=1,2;$ $\ell =1,2,3$)---we refer to the analogous discussion
in \cite{CP2019}.

\bigskip

\section{Outlook}

In this final \textbf{Section 3} we outline tersely possible future
developments of the findings reported above.

The findings reported in this paper extend to evolutions in \textit{%
discrete-time} only some of the findings for evolutions in \textit{%
continuous-time} reported in \cite{CP2018} and \cite{CP2019} and tersely
reviewed above (in \textbf{Section 1}). It is therefore quite natural to
envisage an extension from the \textit{continuous-time} context to the
\textit{discrete-time} context of other results reported in \cite{CP2018}
and \cite{CP2019}.

A different research line might be directed towards \textit{applications} of
the \textit{solvable} discrete-time evolution equation (\ref{2zntilde}),
including its generalization via an assigned shift of the dependent
variables $z_{n}\left( \ell \right) $, say
\end{subequations}
\begin{subequations}
\begin{equation}
z_{n}\left( \ell \right) =w_{n}\left( \ell \right) +f_{n}\left( \ell \right)
~,~~~n=1,2~,  \label{zwfel}
\end{equation}%
with $f_{n}\left( \ell \right) $ two \textit{arbitrarily assigned} functions
of the discrete-time $\ell ,$ implying that the new dependent-variable $%
w_{n}\left( \ell \right) $ satisfies the (of course still \textit{solvable})
discrete-time evolution equation%
\begin{eqnarray}
\tilde{w}_{n} &=&a_{n1}\left( w_{1}\right) ^{2}+a_{n2}\left( w_{2}\right)
^{2}+a_{n3}w_{1}w_{2}+g_{n1}w_{1}+g_{n2}w_{2}+h_{n}~,  \notag \\
g_{n1}\left( \ell \right) &\equiv &2a_{n1}f_{1}\left( \ell \right)
+a_{n3}f_{2}\left( \ell \right) ~,~~~g_{n2}\left( \ell \right) \equiv
2a_{n2}f_{2}\left( \ell \right) +a_{n3}f_{1}\left( \ell \right) ~,  \notag \\
h_{n}\left( \ell \right) &\equiv &a_{n1}\left[ f_{1}\left( \ell \right) %
\right] ^{2}+a_{n2}\left[ f_{2}\left( \ell \right) \right]
^{2}+a_{n3}f_{1}\left( \ell \right) f_{2}\left( \ell \right) -f_{n}\left(
\ell +1\right) ~,  \notag \\
n &=&1,2~.  \label{wntilde}
\end{eqnarray}

\bigskip

\section{Acknowledgements}

FP likes to thank the Physics Department of the University of Rome\ "La
Sapienza" for the hospitality from April to November 2018 (during her
sabbatical), when the results reported in this paper were obtained.

\bigskip

\section{Appendix A: A useful class of \textit{solvable} systems of $2$
nonlinear \textit{discrete-time} evolution equations for the $2$ variables $%
y_{m}\left( \ell \right) $}

The system (\ref{21y12tilde}) of \textit{discrete-time }evolution equations
discussed in this \textbf{Appendix A} is too simple to justify considering
its solution as a \textit{new} finding; its solution is reported here
because of its role in solving the novel, more interesting, \textit{%
discrete-time} model discussed above. The solution of the initial-value
problem of the first of the $2$ \textit{discrete-time} evolution equations (%
\ref{21y12tilde}),
\end{subequations}
\begin{subequations}
\begin{equation}
\tilde{y}_{1}=\alpha \left( y_{1}\right) ^{2}~,
\end{equation}%
is an easy task:%
\begin{equation}
y_{1}\left( \ell \right) =\alpha ^{-1}\left[ \alpha y_{1}\left( 0\right) %
\right] ^{2^{\ell }}~,~~~\ell =0,1,2,...~.  \label{y1el}
\end{equation}

To solve the initial-value problem for the second of the $2$ \textit{%
discrete-time} evolution equations (\ref{21y12tilde}),
\end{subequations}
\begin{equation}
\tilde{y}_{2}=\beta ^{2}\left( y_{1}\right) ^{2}y_{2}+\gamma \left(
y_{1}\right) ^{4}~,  \label{y2tilde}
\end{equation}%
it is convenient to introduce the \textit{ansatz}
\begin{subequations}
\label{y2Y}
\begin{equation}
y_{2}\left( \ell \right) =\left\{ \beta ^{2\ell }\tprod\limits_{s=0}^{\ell
-1}\left[ y_{1}\left( s\right) \right] ^{2}\right\} Y\left( \ell \right) ~,
\label{AYy2a}
\end{equation}%
implying%
\begin{equation}
Y\left( 0\right) =y_{2}\left( 0\right) ~.  \label{Yy2at0}
\end{equation}

\textbf{Remark A-1}. We always use the standard convention according to
which, if $s_{1}>s_{2}$,
\end{subequations}
\begin{equation}
\sum_{s=s_{1}}^{s_{2}}f\left( s\right)
=0~,~~~\tprod\limits_{s=s_{1}}^{s_{2}}f\left( s\right) =1
\end{equation}%
for \textit{any arbitrary} function $f\left( s\right) $ of the \textit{%
discrete-time} variable $s$. $\blacksquare $\hspace{0in}

It is then easily seen that (\ref{y2tilde}) implies
\begin{subequations}
\label{AYF}
\begin{equation}
Y\left( \ell +1\right) =Y\left( \ell \right) +F\left( \ell \right) ~,
\label{AYFa}
\end{equation}%
with%
\begin{equation}
F\left( \ell \right) =\gamma \left[ y_{1}\left( \ell \right) \right]
^{4}\beta ^{-2\left( \ell +1\right) }\tprod\limits_{s=0}^{\ell }\left[
y_{1}\left( s\right) \right] ^{-2}~,
\end{equation}%
implying, via (\ref{y1el}),%
\begin{equation}
F\left( \ell \right) =\gamma \alpha ^{-4}\left( \alpha /\beta \right)
^{2\left( \ell +1\right) }\left[ \alpha y_{1}\left( 0\right) \right] ^{2}~.
\label{AYFc}
\end{equation}

Hence in conclusion, from (\ref{AYF}) with (\ref{Yy2at0}),
\end{subequations}
\begin{subequations}
\begin{equation}
Y\left( \ell \right) =y_{2}\left( 0\right) +\gamma \left( \alpha \beta
\right) ^{-2}\left[ \frac{\left( \alpha /\beta \right) ^{2\ell }-1}{\left(
\alpha /\beta \right) ^{2}-1}\right] \left[ \alpha y_{1}\left( 0\right) %
\right] ^{2}~,
\end{equation}%
hence, via (\ref{AYy2a}),%
\begin{eqnarray}
&&y_{2}\left( \ell \right) =\left( \beta /\alpha \right) ^{2\ell }\left[
\alpha y_{1}\left( 0\right) \right] ^{2^{\ell +1}-2}\left\{ y_{2}\left(
0\right) \right.  \notag \\
&&\left. +\gamma \left( \alpha \beta \right) ^{-2}\left[ \frac{\left( \alpha
/\beta \right) ^{2\ell }-1}{\left( \alpha /\beta \right) ^{2}-1}\right] %
\left[ \alpha y_{1}\left( 0\right) \right] ^{2}\right\} ~.  \label{y2el}
\end{eqnarray}

The $2$ formulas (\ref{y1el}) and (\ref{y2el}) provide the \textit{explicit}
solution of the initial-values problem of the discrete-time evolution (\ref%
{21y12tilde}). Of course in (\ref{y1el}), to make this solution applicable
to the final findings reported in \textbf{Section 2}, the assignment $\gamma
=\left( \alpha ^{2}-\beta ^{2}\right) /4$ must be made, see (\ref{21gamma}).

\end{subequations}

\end{document}